\pdfoutput=1
\documentclass[11pt,twoside,a4paper,cmspaper,final,collab]{cms-tdr}
\def\svnVersion{e34a597}\def\svnDate{2026/06/29}\def\cmsCernNoTag{CERN-EP-2026-025}\def\cmsCernDate{\today}\def\cmsMessage{Published in Physics Letters B as \href{http://dx.doi.org/10.1016/j.physletb.2026.140671}{\doi{10.1016/j.physletb.2026.140671}.}}
\begin{document}\cmsNoteHeader{EXO-25-007}

\newcommand{\dark}{\ensuremath{\text{D}}\xspace}
\newcommand{\Pphi}{\ensuremath{\phi_{\dark}}\xspace}
\newcommand{\mPhi}{\ensuremath{m_{\Pphi}}\xspace}
\DeclareRobustCommand{\PAprime}{{\HepParticle{A}{}{\prime}}\Xspace}
\newcommand{\mAprime}{\ensuremath{m_{\PAprime}}\xspace}
\newcommand{\Pj}{\ensuremath{\text{j}}\xspace}
\newcommand{\TD}{\ensuremath{T_{\dark}}\xspace}
\newcommand{\lamtwo}{\ensuremath{\lambda_{2}}\xspace}
\newcommand{\lamthr}{\ensuremath{\lambda_{3}}\xspace}
\newcommand{\pTleadjet}{\ensuremath{\pt^{\Pj_{1}}}\xspace}
\newcommand{\mll}{\ensuremath{m_{\Pell\Pell}}\xspace}
\newcommand{\BR}{\ensuremath{\mathcal{B}}\xspace}
\newcommand{\sigmaProd}{\ensuremath{\sigma^{\text{prod}}}\xspace}
\newcommand{\BRHSUEP}{\ensuremath{\BR(\PH\to\text{SUEP})}}
\newcommand{\sigmatimesBR}{\ensuremath{\sigmaProd\BRHSUEP}\xspace}
\newcommand{\sigmatimesBRoverSM}{\ensuremath{\sigmatimesBR/\sigmaSM}\xspace}
\newcommand{\sigmaSM}{\ensuremath{\sigmaProd_{\text{SM}}}\xspace}
\newcommand{\pipi}{\ensuremath{\PGpp\PGpm}\xspace}
\newcommand{\cmslumi}{\ensuremath{138\fbinv}\xspace}
\newcommand{\nSUEPConstituents}{\ensuremath{n_{\text{constituent}}^{\text{SUEP}}}\xspace}
\newcommand{\boostedSphericity}{\ensuremath{S^{\text{SUEP}}_{\text{boosted}}}\xspace}
\newcommand{\abseta}{\ensuremath{\abs{\eta}}\xspace}
\newcommand{\MADANALYSIS} {\textsc{MadAnalysis}\xspace}
\newcommand{\DELPHES} {\textsc{Delphes}\xspace}
\newcommand{\COMBINE} {\textsc{Combine}\xspace}
\newcommand{\wjets}{{{\PW}+jets}\xspace}
\newcommand{\WH}{\ensuremath{\PW\PH}\xspace}
\newcommand{\ZH}{\ensuremath{\PZ\PH}\xspace}
\newcommand{\VH}{{{\WH}+{\ZH}}\xspace}
\newcommand{\NAa}{\ensuremath{N_\text{A}}\xspace}
\newcommand{\NB}{\ensuremath{N_\text{B}}\xspace}
\newcommand{\NC}{\ensuremath{N_\text{C}}\xspace}
\newcommand{\ND}{\ensuremath{N_\text{D}}\xspace}
\newcommand{\NE}{\ensuremath{N_\text{E}}\xspace}
\newcommand{\NF}{\ensuremath{N_\text{F}}\xspace}
\newcommand{\NGg}{\ensuremath{N_\text{G}}\xspace}
\newcommand{\NH}{\ensuremath{N_\text{H}}\xspace}
\newcommand{\NI}{\ensuremath{N_\text{I}}\xspace}
\newlength\cmsFigWidth
\ifthenelse{\boolean{cms@external}}{\setlength\cmsFigWidth{0.49\textwidth}}{\setlength\cmsFigWidth{0.8\textwidth}}

\cmsNoteHeader{EXO-25-007}

\title{Search for soft unclustered energy patterns produced in association with a \texorpdfstring{\PW}{W} or \texorpdfstring{\PZ}{Z} boson in proton-proton collisions at \texorpdfstring{$\sqrt{s} = 13\TeV$}{sqrt(s) = 13 TeV}}

\date{\today}

\abstract{A search for a Higgs boson produced in association with a \PW or \PZ boson and decaying via a soft unclustered energy pattern (SUEP) is presented. The analysis is based on proton-proton collision data corresponding to an integrated luminosity of \cmslumi collected between 2016 and 2018 at the LHC. Final states with a leptonic \PW or \PZ boson decay associated with a high multiplicity of low-momentum charged particles are explored for the first time. The results show no significant excess over the standard model background expectation. Limits are set on the production cross section of a Higgs boson that decays to a SUEP, for a range of parameters of the SUEP model. Material is provided to facilitate further interpretation of the results.}

\hypersetup{
pdfauthor={CMS Collaboration},
pdftitle={Search for soft unclustered energy patterns in association with a W or Z boson in proton-proton collisions at sqrt(s) = 13 TeV},
pdfsubject={CMS},
pdfkeywords={CMS, EXO, dark showers, SUEP, hidden valley}}

\maketitle 

\section{Introduction}\label{sec:introduction}
The extensive searches at the CERN LHC have probed numerous final states without direct evidence for physics beyond the standard model (BSM). As many minimal BSM models have now been ruled out, more unusual signatures that may have evaded previous searches are being explored as possible explanations for questions unanswered by the standard model (SM). A well-motivated class of BSM models are hidden-valley (HV) models~\cite{Cvetic:2002qa,Chacko:2005pe,Strassler:2006im,Abel:2008ai,Hur:2007uz,Bai:2013xga}, which can provide solutions to the dark matter puzzle~\cite{Bai:2013xga} and the hierarchy problem~\cite{Chacko:2005pe}. The HV models extend the SM with a non-Abelian gauge group, dark quantum chromodynamics (QCD), and predict unconventional signatures including emerging~\cite{Schwaller:2015gea} or semivisible~\cite{Cohen:2015toa} jets. If the product of the running dark QCD coupling and the number of dark QCD colors ('t Hooft coupling~\cite{tHooft:1973alw}) exceeds unity for a broad range of energies above the confinement scale, collinear emission is no longer enhanced. This dark QCD showering produces low-momentum particles spherically distributed in the rest frame of the initiating particle~\cite{Polchinski:2002jw,Strassler:2006im,Hofman:2008ar}, a signature known as a soft unclustered energy pattern (SUEP)~\cite{Strassler:2008bv,Knapen:2016hky}. In the considered phenomenological model~\cite{Knapen:2016hky}, a heavy scalar mediator produces a SUEP shower that is composed of multiple, identical dark mesons (\Pphi) that decay to dark photons (\PAprime), which in turn decay to visible SM particles, producing a signature of numerous low-momentum (soft) SM particles spread broadly in the $\eta$-$\phi$ space.

\begin{figure}[htbp]
    \centering
    \includegraphics[width=.48\textwidth]{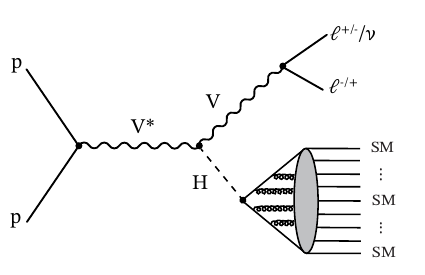}
        \caption{Schematic diagram of the 125\GeV Higgs boson decaying to a SUEP shower, produced in association with a leptonically decaying \PV boson.}
\label{fig:feynman}
\end{figure}

The CMS Collaboration has previously searched for SUEPs from the decays of heavy scalar mediators produced through gluon fusion~\cite{CMS:2024nca}. A subsequent search by the ATLAS Collaboration targeted models with enhanced muon content in the final state~\cite{ATLAS:2026fso}. The case in which the mediator is the Higgs boson (\PH) is especially interesting, as it can be produced in multiple ways without increasing the complexity of the dark sector. Production modes other than gluon fusion enable more efficient triggering strategies~\cite{Knapen:2016hky}. This Letter investigates the hypothesis that the scalar is the 125\GeV Higgs boson produced in association with a massive vector boson, \PW or \PZ (\PV), which decays to leptons, as depicted in Fig.~\ref{fig:feynman}. The analysis uses triggers based on these final-state electrons and muons. This requirement significantly reduces the multijet background compared to the previous CMS search exploring gluon fusion production~\cite{CMS:2024nca}, although it introduces the challenge of separating the signal from the Drell--Yan (DY), \wjets, and top quark-antiquark pair (\ttbar) backgrounds. 

\section{The CMS detector}\label{sec:CMS}
The CMS apparatus~\cite{CMS:2008xjf} is a multipurpose, nearly hermetic detector, designed to trigger on~\cite{CMS:2016ngn,CMS:2020cmk,CMS:2024aqx} and identify electrons, muons, photons, and (charged and neutral) hadrons~\cite{CMS:2020uim,CMS:2018rym,CMS:2014pgm}. A global ``particle-flow'' (PF) algorithm~\cite{Sirunyan:2017ulk} aims to reconstruct all individual particles in an event, combining information provided by the all-silicon tracker and by the crystal electromagnetic and brass-scintillator hadron calorimeters, operating inside a 3.8\unit{T} superconducting solenoid, with data from the gas-ionization muon detectors embedded in the flux-return yoke outside the solenoid. The primary vertex (PV) is taken to be the vertex corresponding to the hardest scattering in the event, evaluated using tracking information alone (Ref.~\cite{CMS-TDR-15-02}, Section 9.4.1).

\section{Signal and background simulation}\label{sec:samples}
Simulated samples of the $\PV\PH$ signal process are used to optimize the analysis procedure, estimate the selection efficiency, and interpret the results. The $\PV\PH$ production process is simulated at leading order (LO) with \PYTHIA 8.240~\cite{Sjostrand:2014zea} using the BSM Higgs boson library, with the Higgs boson mass set to 125\GeV~\cite{Sjostrand:2014zea}. In simulation, the Higgs boson decay produces a SUEP shower composed of multiple dark mesons. The production of dark mesons in the SUEP shower is simulated using a \PYTHIA plugin based on Ref.~\cite{Knapen:2016hky}, which models the SUEP shower as an isotropic emission of dark mesons with Boltzmann-distributed transverse momentum \pt. Each dark meson decays to a pair of dark photons through an effective $\Pphi \PAprime^{\mu\nu} \PAprime_{\mu\nu}$ interaction. The dark photons decay to visible SM particles through kinetic mixing with the SM photon~\cite{Buschmann:2015awa,Graham:2021ggy}. Signal samples are normalized to next-to-next-to-LO (NNLO) cross section predictions for Higgs boson production~\cite{LHCHiggsCrossSectionWorkingGroup:2016ypw}. The parameters probed in the model are the mass of the dark meson \mPhi, the Boltzmann temperature \TD, and the mass of the dark photon \mAprime. Two \mAprime values correspond to dominantly leptonic ($\mAprime=0.5\GeV$, $\PAprime \to \EE, \MM,~\text{and}~\pipi$, with branching fractions $\BR = 40, 40,~\text{and}~20\%$, respectively) or dominantly hadronic ($\mAprime=0.7\GeV$, $\PAprime \to \EE, \MM,~\text{and}~\pipi$, with $\BR = 15, 15,~\text{and}~70\%$, respectively) decays. At $\mAprime=1.02\GeV$, the dark photon mixes with the $\PGf(1020)$ meson and decays almost always to kaons. While this model is fairly unique because of the narrow width of the $\PGf(1020)$ meson, we include a model with $\mAprime=1\GeV$ and 100\% branching fraction to \pipi as a generic proxy for a fully hadronic SUEP decay, as used in the previous CMS search~\cite{CMS:2024nca}. The \mPhi values are varied between 1 and 8\GeV with \TD between $\mPhi/4$ and $4\mPhi$, following the constraints discussed in Ref.~\cite{Barron:2021btf}. 

SM simulated samples are used in the validation of the background estimation procedure. Multijet samples are produced at LO with \PYTHIA. The \MGvATNLO 2.6.5~\cite{Alwall:2014hca} program is used for the simulation of the DY, \PW, single top quark, and $\ttbar+\PV$ processes at next-to-LO (NLO). The DY and \PW ($\ttbar+\PV$) samples are produced with up to two (one) additional partons at the matrix element level and the FxFx merging scheme is used~\cite{Frederix:2012ps}. The {\POWHEG} \textsc{Box} v2~\cite{Nason:2004rx,Frixione:2007vw,Alioli:2010xd,Alioli:2011as,Nason:2013ydw} is used for the simulation of the \ttbar and diboson processes at NLO. 

All signal and background samples are produced using the NNPDF 3.1 parton distribution function (PDF) set~\cite{Ball:2017nwa} at next-to-NLO. The underlying event, parton showering, and additional proton-proton ($\Pp\Pp$) interactions in the same or adjacent bunch crossings (pileup) are simulated with \PYTHIA using the CP5 tune~\cite{Sirunyan:2019dfx}. A realistic simulation of the response of the CMS detector is performed with the \GEANTfour package~\cite{Agostinelli:2002hh}.

\section{Object reconstruction and identification}\label{sec:objects}
Charged leptons are selected to target those produced from the decay of a \PV boson rather than in the SUEP shower. Selected electrons must satisfy $\pt > 15\GeV$ and pseudorapidity $\abseta < 2.5$. A multivariate analysis (MVA) discriminant that combines information from the reconstructed electron and its isolation is used to reject poorly reconstructed electrons and those coming from hadron decays~\cite{CMS:2020uim}. Restrictions on the transverse and longitudinal impact parameters of the electron tracks are imposed to ensure consistency with the PV. Two different selections based on the MVA discriminant are used, corresponding to selection efficiencies of roughly 90\% (loose) and 80\% (tight). Selected muons must have $\pt > 10\GeV$ and $\abseta < 2.4$. Additional identification criteria are used to define loose (tight) muons~\cite{CMS:2018rym}, which are also required to pass loose (very tight) relative PF isolation requirements, corresponding to a selection efficiency of 98 (91)\%~\cite{CMS:2018rym}. Events with exactly one (two) selected loose electrons or muons are assigned to the \WH (\ZH) channel.

{\tolerance=800 Hadronic jets are used to apply different event selection criteria. After rejecting charged hadrons that do not originate from the PV, the anti-\kt algorithm~\cite{Cacciari:2008gp,Cacciari:2011ma} with a distance parameter of $R = 0.4$ is used to cluster the reconstructed particles into jets; subsequently, neutral pileup contributions are subtracted~\cite{CMS:2016lmd}. Jets must have $\pt > 30\GeV$, $\abseta < 2.5$, and pass tight identification criteria designed to reduce noise effects from the calorimeter systems with a 99\% efficiency for genuine jets~\cite{CMS-PAS-JME-16-003}. The reconstructed jet energy is corrected to better match the energy of the originating parton~\cite{CMS:2016lmd}. Jets produced by the hadronization of \PQb quarks (\PQb jets) are identified, or ``tagged'', using the loose and tight working points of the \textsc{DeepJet} algorithm~\cite{BTV-16-002,Bols:2020bkb,CMS-DP-2023-005}, which, respectively, have misidentification rates of 10 and 0.1\% for jets originating from light quarks or gluons, and efficiencies of 90 and 55\% for genuine \PQb quark jets. Any jet (\Pj) that overlaps with a selected electron or muon (\Pell), namely that has $\Delta R(\Pell, \Pj) = \sqrt{(\Delta\phi)^2+(\Delta\eta)^2} < 0.4$, is discarded. The missing transverse momentum vector \ptvecmiss, with magnitude \ptmiss, is computed from the PF candidates weighted by their probability to originate from the PV~\cite{CMS:2019ctu}, as predicted by the pileup-per-particle identification algorithm~\cite{Bertolini:2014bba}.\par}

To build reconstructed SUEP objects, all charged PF candidates, independently of whether they are clustered within a hadronic jet, that originate from the PV with $\pt > 1.0\GeV$, $\abseta < 2.5$ are clustered using the anti-\kt algorithm with a distance parameter of $R = 1.5$, consistent with the expected size in the $\eta$--$\phi$ plane of a SUEP recoiling against a \PV boson. The cluster with the highest \pt in each event is defined to be the reconstructed SUEP candidate and is required to have $\pt > 60\GeV$. The clustering omits PF candidates overlapping with a selected lepton by $\Delta R(\Pell, \text{PF candidate}) < 0.4$ to avoid contamination from emissions of the \PW and \PZ boson decays. 

The associated production of a \PH boson and a photon has a negligible cross section. Thus, photons are used to define signal-free independent validation regions where they act as a proxy for the \PW or \PZ boson. Photons must satisfy $\pt > 60\GeV$ and $\abseta < 2.5$. Identification and isolation criteria are applied to all photons, based on an MVA discriminant similar to that used for electrons~\cite{CMS:2020uim}. The selection efficiency for photons is ${\approx}90\%$. 

\section{Event selection}\label{sec:selection}
We analyze a data sample of $\Pp\Pp$ collisions at $\sqrt{s}=13\TeV$ collected by the CMS experiment between 2016 and 2018, corresponding to an integrated luminosity of \cmslumi. Events are selected with several single and double electron and muon triggers, chosen to maximize the trigger acceptance. Minimum \pt trigger thresholds vary from 27 to 32\GeV (24 to 27\GeV) for single-electron (single-muon) triggers. For double-lepton paths they correspond to 23 (17)\GeV and 12 (8)\GeV for the leading and subleading electron (muon). Additional high-\pt triggers with relaxed criteria target electrons (muons) with a \pt threshold of 115 (50)\GeV. These triggers increase the signal acceptance inclusively by roughly 1\%, with the effects of up to 7\% (2\%) for electrons (muons) with $\pt > 115 \GeV$ ($\pt > 50 \GeV$).

To account for the different sources of SM background in each channel, different selection criteria and background estimation strategies are employed to define signal regions (SRs) and control regions (CRs).

The \WH channel selection requires events to pass single-lepton triggers and to have exactly one selected tight lepton and a SUEP candidate. Muons (electrons) must have $\pt > 30$ (35)\GeV to account for the more stringent criteria of the single-lepton triggers. Events are required to have $\ptmiss > 30\GeV$. The \PW boson is reconstructed by combining the transverse momentum vector of the selected lepton and the \ptvecmiss. Events are required to have $\pt^{\PW} > 60\GeV$ and $30 < \mT^{\PW} < 130\GeV$, where $\mT^{\PW} = \sqrt{\smash[b]{2 \pt^{\Pell} \ptmiss [1 - \cos{\Delta\phi (\ell, \ptvecmiss)}]}}$ is the transverse mass of the lepton-\ptmiss system. To select the desired final state of a SUEP candidate recoiling against a \PW boson, the ratio of the two objects' momenta is constrained to be $\pt^{\PW}/\pt^{\text{SUEP}} < 3$, and the differences in azimuth between the SUEP candidate and each of the \PW boson, the lepton, and the \ptvecmiss must be greater than 1.5 radians. The sphericity of the SUEP candidate in its rest frame is defined using the second and third largest eigenvalues (\lamtwo, \lamthr) of the infrared- and collinear-safe generalized sphericity tensor: $\boostedSphericity \equiv \frac{3}{2} ( \lamtwo + \lamthr )$. $\boostedSphericity$ is expected to be lower in SM production where isotropical patterns are suppressed~\cite{Cesarotti:2020ngq}. A requirement on $\boostedSphericity > 0.3$ is applied. Events with two loose or one tight \PQb-tagged jets are rejected to reduce the \ttbar background. This selection defines the \WH SR (SR-WH). Simulation predicts the SM yield in SR-WH to be 88\% from \wjets production, with roughly equal subdominant contributions from \ttbar (3\%), DY (3\%), and multijet production (4\%), and residual contributions from diboson and single top quark processes (2\%).

The \ZH channel selection requires events to pass single- or double-lepton triggers and to have exactly two loose leptons of same flavor and opposite charge, plus a SUEP candidate overlapping with at least one jet. The \pt of the leading overlapping jet (\pTleadjet) is expected to be higher for SM processes, where the entire jet fits into a cone with $R=0.4$, than for SUEPs, where only a subset of the SUEP constituents will be included in the cluster. The leading lepton must have $\pt > 25\GeV$, based on the double-lepton trigger criteria. The \PZ boson is reconstructed by adding the four-momenta of the two leptons. The reconstructed \PZ boson must satisfy $\pt^{\PZ} > 25\GeV$ and $60 < \mll < 120\GeV$. The \ZH SR (SR-ZH) is defined by applying these criteria and rejecting events with at least one loose \PQb-tagged jet to reduce the \ttbar background. Two CRs are defined by requiring $\mll > 120\GeV$: CR-DY, which maintains the \PQb-tagged jet veto, and CR-TT, which instead requires a single loose \PQb-tagged jet. Simulation predicts the SM yield in SR-ZH to be mostly from DY production (95\%), with subdominant contributions from diboson (4\%) and \ttbar (1\%) production. In CR-TT, simulation predicts a majority events from the \ttbar background (80\%), with significant contributions from DY (16\%), single top quark (2\%), and diboson (2\%) processes. In CR-DY, the dominant process is DY (79\%), followed by diboson processes (11\%), \ttbar (9\%), and single top quark processes (1\%).

Photon triggers with \pt thresholds ranging from 50 to 200\GeV are used to define signal-free regions to validate the background estimation procedure. These triggers collect only part of the data, as described in Section~\ref{sec:PR}. For the \WH (\ZH) channel, a photon region PR-WH (PR-ZH) is defined using the same selection criteria as SR-WH (SR-ZH), replacing the requirements for the reconstructed \PW (\PZ) boson with those for the photon, omitting the transverse (dilepton) mass requirement. The PR-WH and PR-ZH regions are used for validation and are not included in the final fit to the data.

\section{Background estimation}\label{sec:backgrounds}

Particle multiplicity distributions in simulation are especially sensitive to the parton shower model and to the tracking efficiency uncertainties, which can introduce sizeable variations. Thus, the yields and distributions of the SM background processes are allowed to deviate from the compositions stated in Section~\ref{sec:selection}. They are estimated from data with the extended ``ABCD'' method~\cite{Choi:2019mip}. The extended ABCD method relies on dividing the plane of two variables into nine subregions instead of the usual four of the traditional ABCD method. Compared to the four-region ABCD method, the extended ABCD method incorporates corrections for effects from residual dependence between the two variables at the expense of increased statistical uncertainty. All SM background processes are estimated simultaneously in this method, with no direct use of simulation, which is possible because of the similar distribution across the SR bins for all different SM processes, as shown in simulation. 

The two variables chosen for the extended ABCD method are the multiplicity (\nSUEPConstituents) of the SUEP candidate and a channel-specific variable that distinguishes the signal from the SM processes. In the \WH channel, the second variable is \boostedSphericity, which reflects the expected spherical distribution of the SUEP shower. In the \ZH channel, the reconstructed SUEP is slightly less boosted because of a more inclusive selection. As a consequence, for SM processes, the momentum of the jet recoiling against the vector boson, \pTleadjet, is softer and less correlated with \nSUEPConstituents in the \ZH channel than in the \WH channel. As \pTleadjet offers better sensitivity than \boostedSphericity, we chose to use it as the second variable for the \ZH channel. The bin boundaries of the extended ABCD method are chosen for each SR to reduce nonclosure effects, defined as the difference between the observed and predicted yields, based on simulation. Closure tests were performed to compare the performance of the extended ABCD and the four-region ABCD methods in simulation in SR-WH and SR-ZH. In the closure tests, the extended ABCD method was found to reduce the nonclosure from 25\% (20\%) to 2\% (5\%) for SR-WH (SR-ZH). In simulation, closure tests of the method using the SM processes contributing to each SR, as discussed in Section~\ref{sec:selection}, demonstrate that the method accurately reproduces the simulated background yields. This is further confirmed by the tests performed in data with PR-WH and PR-ZH described in Section~\ref{sec:PR}.

The extended ABCD method is implemented as follows. For each SR or CR, eight signal-depleted regions (SDRs), A through H, are defined by placing different boundaries on the two variables. A signal-enriched region (SER), I, is defined for each SR and the corresponding CRs: $\boostedSphericity > 0.5$ and $\nSUEPConstituents > 30$ in the \WH channel, and $\pTleadjet < 135\GeV$ and $\nSUEPConstituents > 20$ in the \ZH channel. The SERs are divided in bins of \nSUEPConstituents. For each channel, the SDRs and SER are included in a binned Poisson likelihood fit, with the mean of the Poisson distribution in each SDR bin represented as an independent free parameter and of each SER bin obtained from the extended ABCD formula below, using the SDR parameters. In the \WH channel, the shape of the background distribution is obtained from region F. In the \ZH channel, the shape of the distribution is obtained from both regions F and C by applying the extended ABCD method to each \nSUEPConstituents bin. The resulting number of SM events predicted for bin $i$ is:
\begin{equation*}
\NI^{i, \WH} = \NF^i \frac{\NF \NH^2 N_{\mathrm{D}}^2 \NB^2}{\NGg \NC \NAa \NE^4},\,
\NI^{i, \ZH} = \frac{(\NF^i)^2}{\NC^i} \frac{\NH^2 N_{\mathrm{D}}^2 \NB^2}{\NGg \NAa \NE^4},
\end{equation*}
where $\NI^{i, \WH}$ is used for SR-WH and $\NI^{i, \ZH}$ for SR-ZH. For CR-TT and CR-DY, a procedure analogous to that of SR-ZH is used, with each selection having a different set of independent free parameters. In the \ZH channel, including SR-ZH, CR-TT, and CR-DY, signal contributions are only considered in the SER, to avoid instabilities in the fit from the nonlinearity of the background estimation method.
In the SDRs, the background yield is always much larger than the signal yield neglected by this procedure.
It was confirmed in simulation that neglecting these contributions changes the expected upper limit by up to 10\%, predominantly for signal models with high \TD or high \mPhi where larger signal contamination appears in the SDRs.

\subsection{Extended ABCD method validation}\label{sec:PR}

\begin{figure*}[hbt]
    \centering
    \includegraphics[width=.98\textwidth]{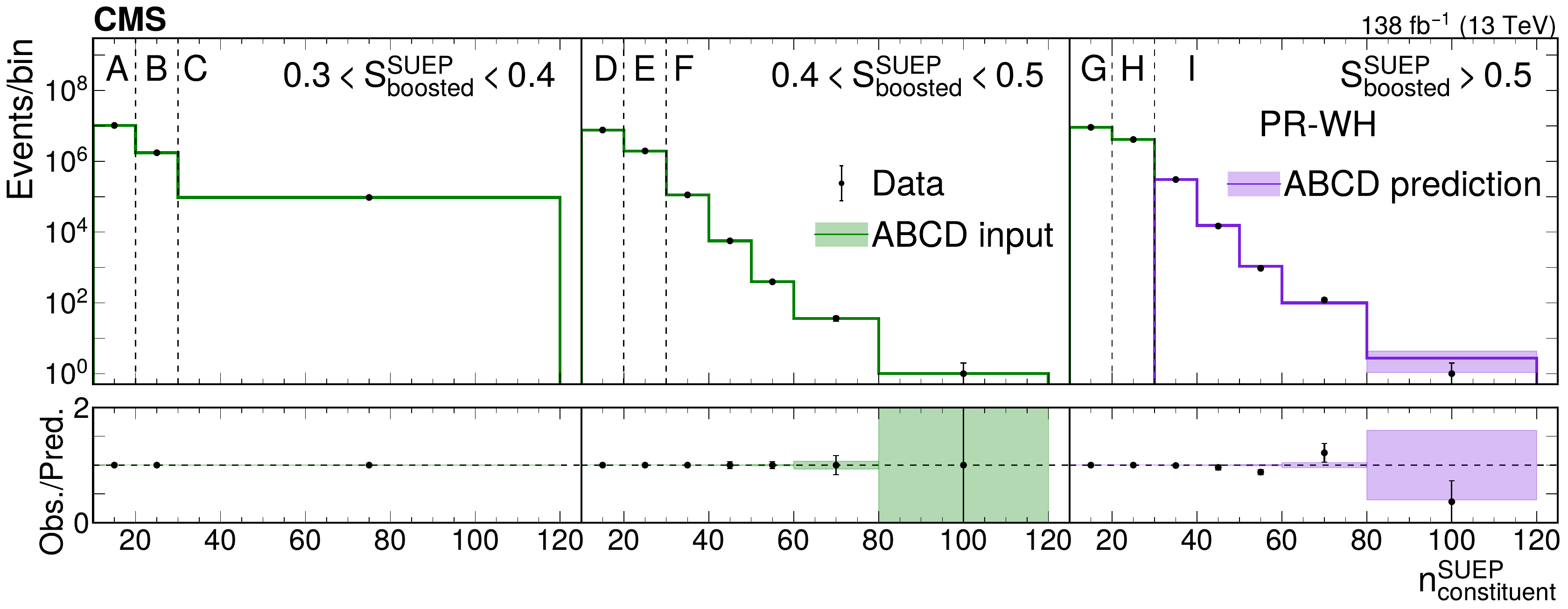}\\
    \includegraphics[width=.98\textwidth]{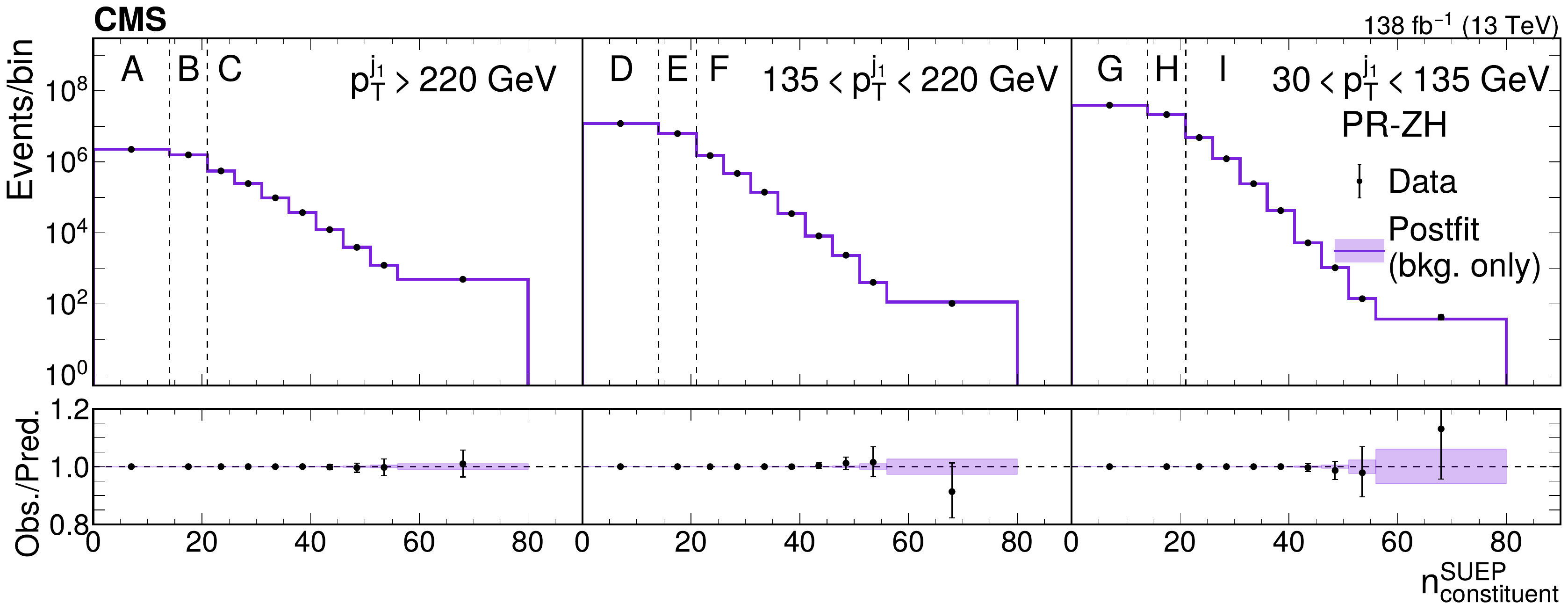}
	\caption{Distributions of \nSUEPConstituents in the extended ABCD subregions A--I in the PR-WH (upper) and PR-ZH (lower) regions, showing the observed data and the background predictions. The prefit (postfit) background distribution is shown for the PR-WH (PR-ZH) channel. The distribution is split into \boostedSphericity (\pTleadjet) bins for the \WH (\ZH) channel. The last \nSUEPConstituents bin includes the overflow events. The shaded bands indicate the uncertainties in the prediction, which are propagated to the ratios in the lower panels.}
    \label{fig:validations}
\end{figure*}

The \WH and \ZH channels use, respectively, PR-WH and PR-ZH with events recorded by photon triggers. Final-state photons are produced much more frequently than either \PW or \PZ bosons at the LHC. Together with the misidentification of QCD jets as final-state photons, this results in the need for additional trigger-level requirements to select photon final states. Thus, PR events are selected with a set of prescaled triggers that collected integrated luminosities between 1 and \cmslumi depending on the photon \pt. This introduces a bias towards higher photon momenta, which would correspondingly bias the kinematic properties of the recoiling SUEP candidate. To correct for this effect, each observed event in both PR-WH and PR-ZH is weighted by the ratio of the full integrated luminosity to that collected by the specific trigger that selected the event. Furthermore, the \PW boson carries nonzero electric charge and therefore has a different initial state than the other production modes. To correct for differences between the \PW and photon signatures, the PR-WH events are weighted such that the observed \pt distribution of the SUEP candidate matches between PR-WH and SR-WH.

In the \WH channel, closure tests for the background estimation method compare the observed data in the PR-WH SER with the prediction obtained from the SDR data. While the resulting distributions are found to be in good agreement, the photon trigger prescaling results in a relatively large statistical uncertainty, higher than the expected uncertainty in SR-WH. Therefore, systematic uncertainties are assigned such that the quadrature sum of the systematic uncertainty and the statistical uncertainty in SR-WH is equal to the PR-WH statistical uncertainty. For each bin of the PR-WH SER, an uncertainty is similarly assigned such that the quadrature sum of this systematic uncertainty, the statistical uncertainty in SR-WH, and the predicted uncertainty in the total yield across all bins of the SER is equal to the relative uncertainty in PR-WH. A 4\% uncertainty is estimated for the overall yield, along with an additional uncertainty uncorrelated across \nSUEPConstituents bins ranging from 0.01 to 100\% for low to high \nSUEPConstituents.  The PR-WH prefit distributions are shown in Fig.~\ref{fig:validations} (upper).

In the \ZH channel, the uncertainties in the background estimation method are determined using simulation in SR-ZH, CR-TT, and CR-DY. The last two are expected to contain additional contributions from the diboson and \ttbar processes, thus serving as a test of the process dependence of the background estimation method. The level of agreement between the background yields predicted directly from simulation and from the extended ABCD method applied to simulated events results in a 20\% uncertainty per bin, uncorrelated across the \nSUEPConstituents bins. A validation fit is performed using the data in PR-ZH, CR-TT, and CR-DY and otherwise following the description of Section~\ref{sec:systematics}. To mimic the presence of signal, signal templates from SR-ZH are used in PR-ZH, only for the purpose of this validation fit. The fits are repeated for all signal hypotheses considered, and there are no significant deviations from the background-only prediction. When simulated signals are injected into the PR-ZH data, the injected signal yields are recovered within one standard deviation. Figure~\ref{fig:validations} (lower) shows the PR-ZH postfit distributions.

\section{Systematic uncertainties and statistical model}\label{sec:systematics}

Several uncertainty sources can introduce variations in the signal \nSUEPConstituents distribution. Correction factors for electron and muon triggering, identification, and isolation criteria are computed with the ``tag-and-probe'' method~\cite{CMS:2018rym}, with uncertainties of up to 1.5\%, depending on the data-taking year and lepton flavor. Uncertainties in the jet energy corrections~\cite{CMS:2016lmd} are propagated through the event selection and account for effects of up to 1 (2)\% for the scale (resolution). The \PQb tagging efficiency is corrected separately for light- and heavy-flavor jets, with uncertainties of up to 3\% depending on the data-taking year~\cite{Bols:2020bkb}. The PF candidate selection efficiency is varied down by 3\%~\cite{CMS:2023lpp} for models with fully hadronic SUEP decays and 4.5\% for signal models with leptons from SUEP decays to account for the lower tracking efficiency for low-\pt electrons, with a resulting effect of up to 30\% in some \nSUEPConstituents bins. The uncertainty in the pileup modeling is assessed by varying the total inelastic $\Pp\Pp$ cross section by 5\%~\cite{CMS:2018mlc}, resulting in an effect of up to 4.5\%. Timing shifts in the hardware trigger~\cite{CMS:2020cmk,CMS:2021yvr} can produce losses of data, for which corrections are applied with an associated uncertainty of up to 1.4\%. The modeling of initial- and final-state radiation is studied by varying the renormalization scale up and down by a factor of 2, independently for the initial- and final-state emissions, with an overall effect of up to 14\%. Uncertainties in the PDFs were found to have effects less than 1\% on the signal distributions in the SR-WH and SR-ZH, so they were not included. The uncertainty in the measured integrated luminosity ranges between 1.2 and 2.3\% for different data-taking years~\cite{CMS:2017sdi,CMS-PAS-LUM-17-004,CMS-PAS-LUM-18-002} and is applied as an overall normalization effect of 1.6\%. Statistical uncertainties from the limited number of events in the data CRs and the simulated event samples are considered as additional sources of uncertainty, uncorrelated across bins. 

\begin{figure*}[!htb]
    \centering
    \includegraphics[width=.98\textwidth]{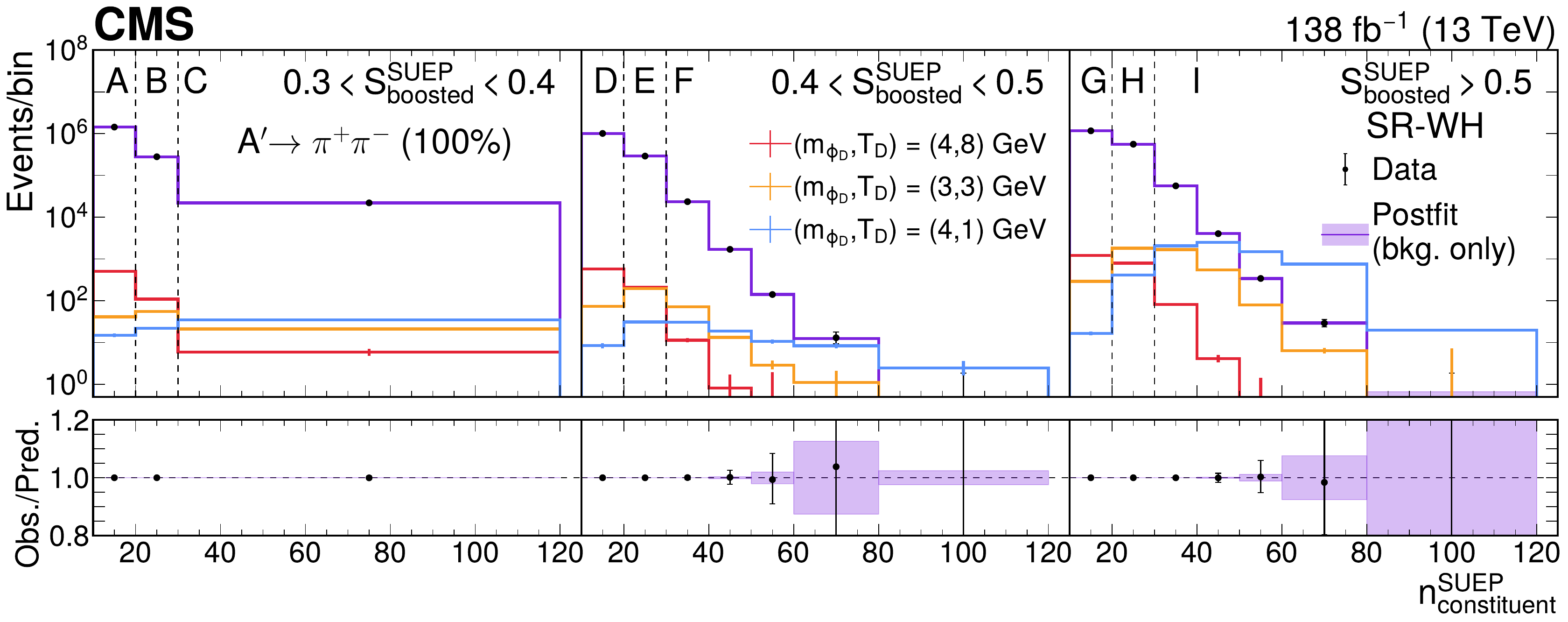}\\
    \includegraphics[width=.98\textwidth]{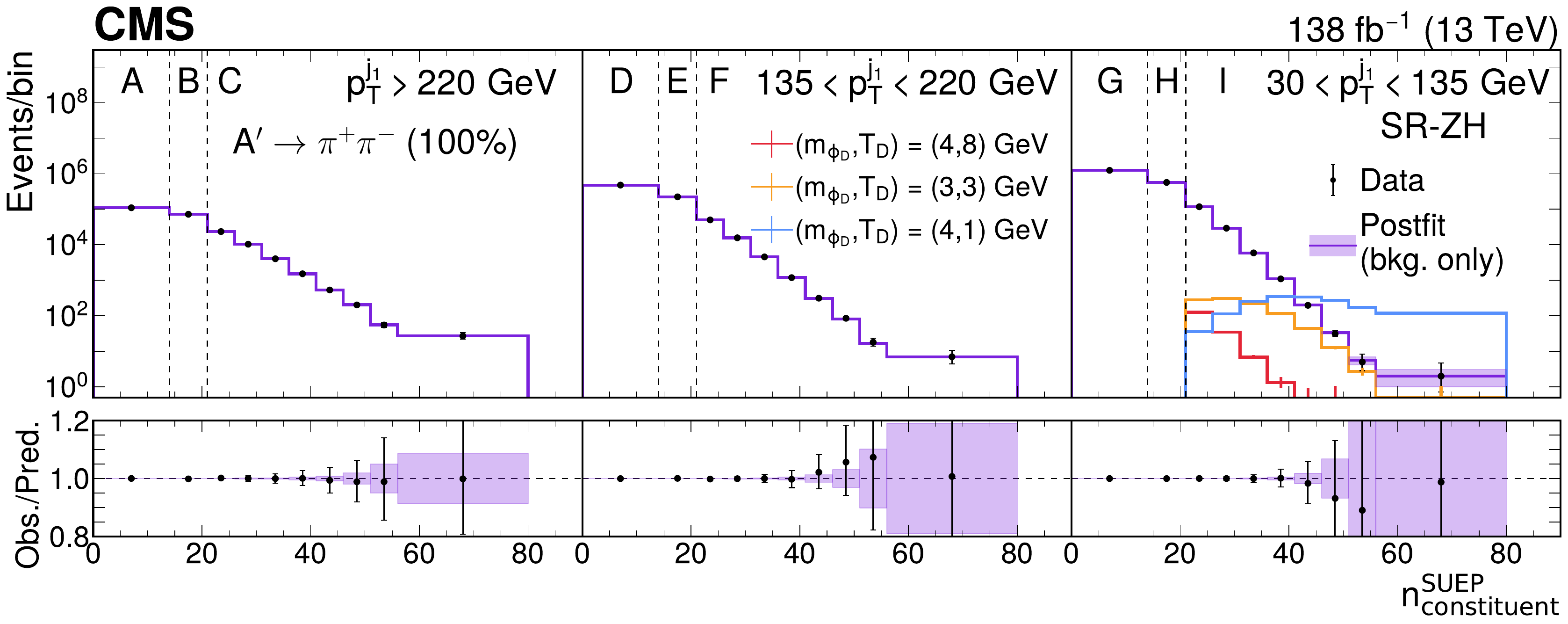}
	\caption{Distributions of \nSUEPConstituents in the extended ABCD subregions A--I in the SR for the \WH (upper) and \ZH (lower) channels, showing the observed data, the predictions from the background-only fit, and several signal hypotheses with fully hadronic decays ($\BR(\PAprime\to\pipi)=100\%$): $\mPhi=4\GeV, \TD=8\GeV$; $\mPhi=3\GeV, \TD=3\GeV$; and $\mPhi=4\GeV, \TD=1\GeV$. The distribution is split into \boostedSphericity (\pTleadjet) bins for the \WH (\ZH) channel. The last \nSUEPConstituents bin includes the overflow events. The shaded bands indicate the uncertainties in the prediction, which are propagated to the ratios in the lower panels. The error bars in the signal distributions correspond to the total prefit uncertainty in the prefit signal distributions.}
    \label{fig:fit}
\end{figure*}

As detailed in Section \ref{sec:backgrounds}, uncertainties in the background estimation procedure are different for the \WH and \ZH channels. In the \WH channel, a 4\% normalization uncertainty affects all the SER bins simultaneously, with a set of decorrelated, independent uncertainties ranging from 0.01 to 100\% affecting each bin of the shape separately. In the \ZH channel, a set of decorrelated uncertainties corresponds to a change in the background yield in each SER bin by 20\%. 

When combining the \WH and \ZH channels, most uncertainty sources are correlated across channels, as they describe effects with common sources in object reconstruction or detector calibration. The exceptions are the uncertainties in the lepton trigger efficiencies, the limited size of the simulated samples, and the background estimation, which are obtained from different data sets.

The extended ABCD subregions from SR-WH, SR-ZH, CR-DY, and CR-TT are included in a binned Poisson likelihood fit, performed using the CMS \COMBINE tool~\cite{CMS:2024onh} and following the description in Section~\ref{sec:backgrounds}. As mentioned in Section~\ref{sec:backgrounds}, the yields in each SDR are represented by free parameters in the fit, meaning that a total of 12 free parameters are included for SR-WH and 22 for each of SR-ZH, CR-DY, and CR-TT. Systematic uncertainties in the SR-ZH background estimation are correlated with those of CR-DY and CR-TT, leading to stronger constraints on the corresponding nuisance parameters. A common parameter represents the product of the \PH boson production cross section (\sigmaProd) times the branching fraction for its decay to a SUEP shower (\BRHSUEP) divided by the SM \VH production cross section of 2.38\unit{pb} (\sigmaSM), denoted as \sigmatimesBRoverSM. The uncertainty sources described above are included as nuisance parameters. This parameter of interest is chosen to be able to provide fair comparisons to previous searches in the gluon fusion channel~\cite{CMS:2024nca} in the case of a \PH boson mediator, by fixing $\sigmaProd = \sigmaSM$ and comparing the branching fraction, while allowing for higher than SM production rates from BSM effects with no restrictions on \sigmaProd.

\section{Results}\label{sec:results}

\begin{figure}[hbt!]
    \centering
    \includegraphics[width=\cmsFigWidth]{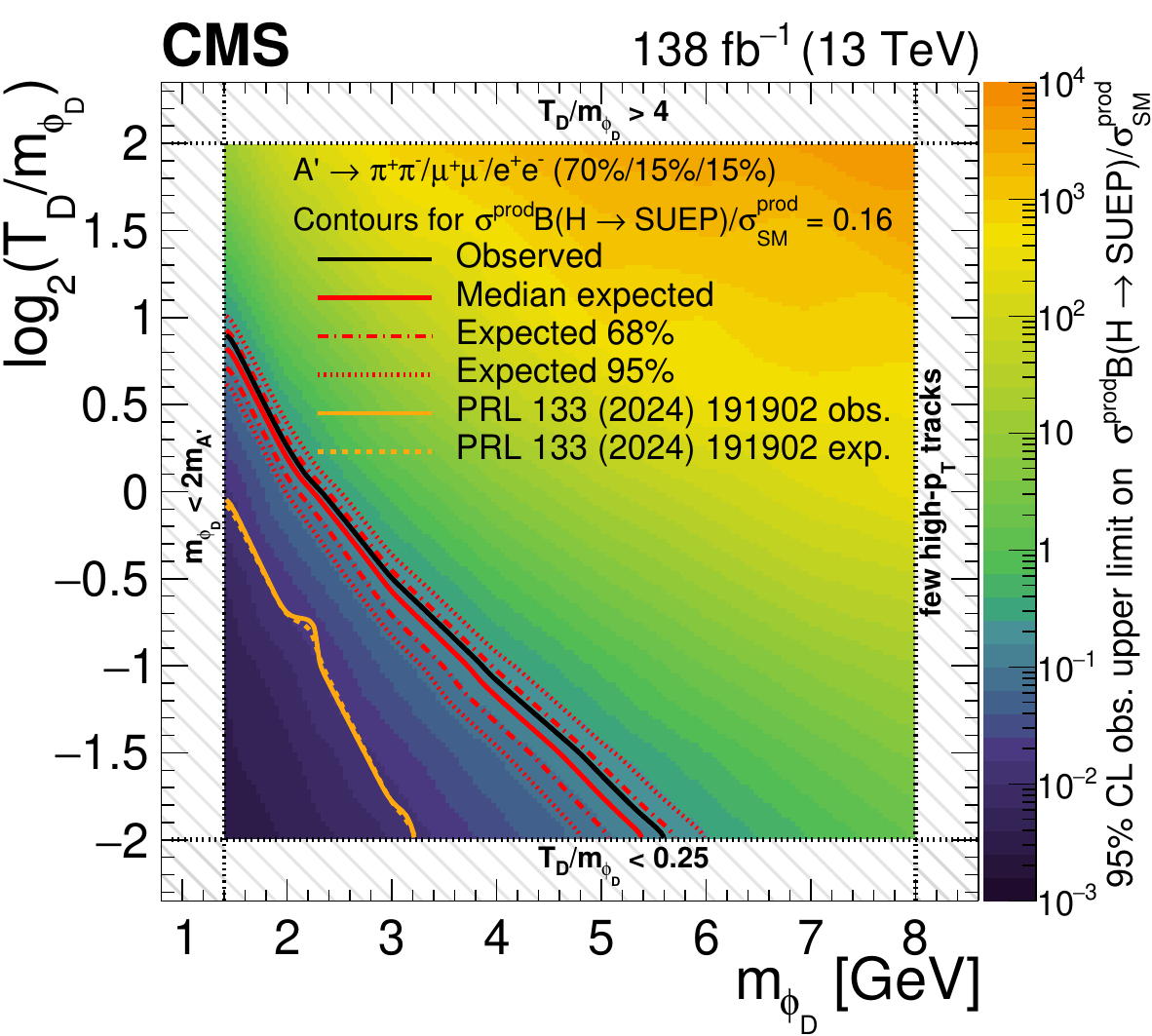}
    \caption{Interpretation of the results for the decay of the 125\GeV Higgs boson mediator to a SUEP, for signal models with dominantly hadronic decays. The observed upper limit from the combination of the \WH and \ZH channels is shown as a function of the signal model parameters \mPhi and $\log_{2}(\TD/\mPhi)$. The contours for each channel are drawn for a benchmark scenario in which the observed or expected upper limit for that channel on \sigmatimesBRoverSM equals 0.16.}
    \label{fig:exclusion}
\end{figure}

\begin{figure}[htb!]
    \centering
    \includegraphics[width=\cmsFigWidth]{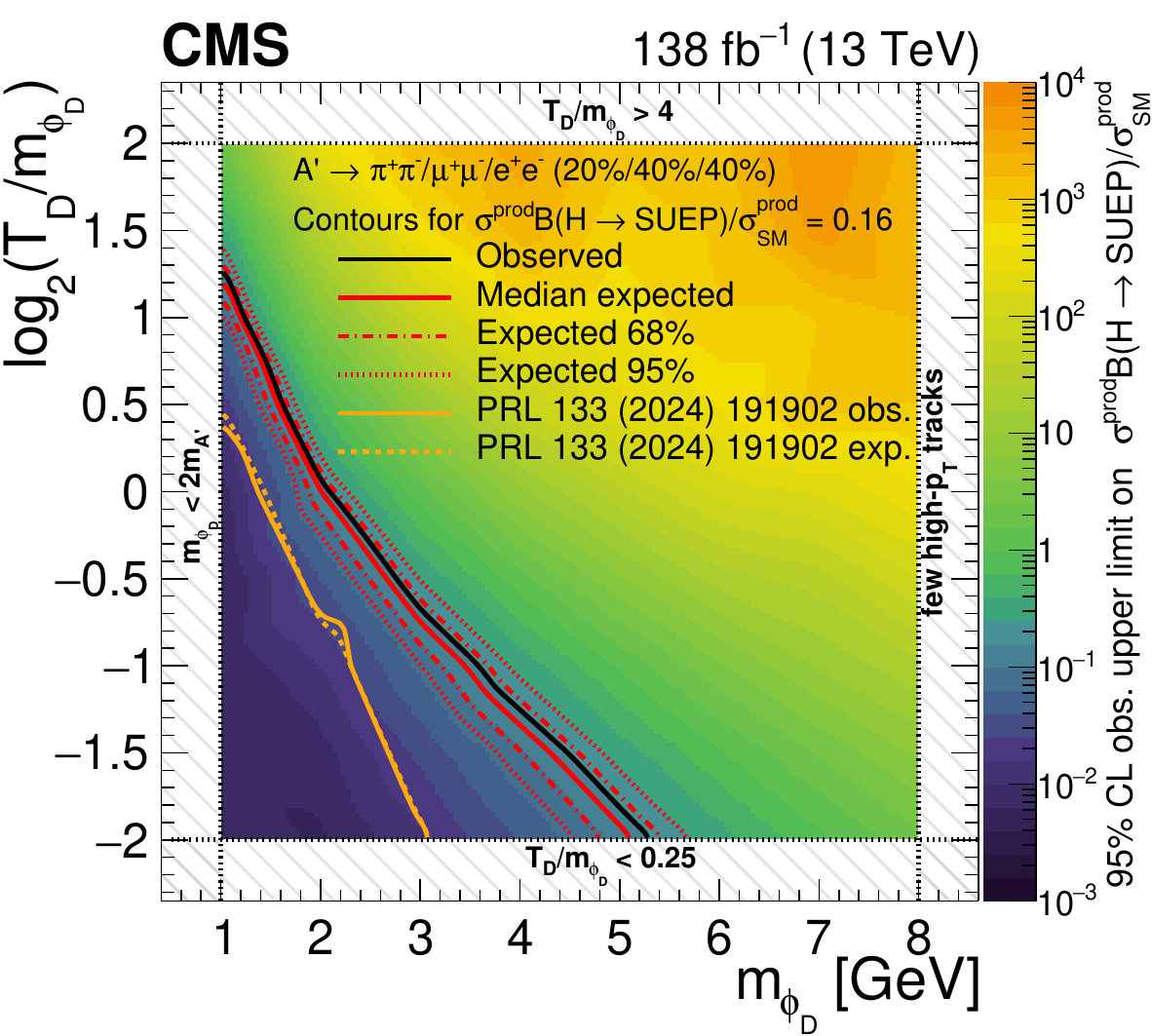}
    \caption{Interpretation of the results for the decay of the 125\GeV Higgs boson mediator to a SUEP, for signal models with dominantly leptonic decays. The observed upper limit from the combination of the \WH and \ZH channels is shown as a function of the signal model parameters \mPhi and $\log_{2}(\TD/\mPhi)$. The contours for each channel are drawn for a benchmark scenario in which the observed or expected upper limit for that channel on \sigmatimesBRoverSM equals 0.16.}
    \label{fig:exclusion_lep}
\end{figure}

\begin{figure}[htb!]
    \centering
    \includegraphics[width=\cmsFigWidth]{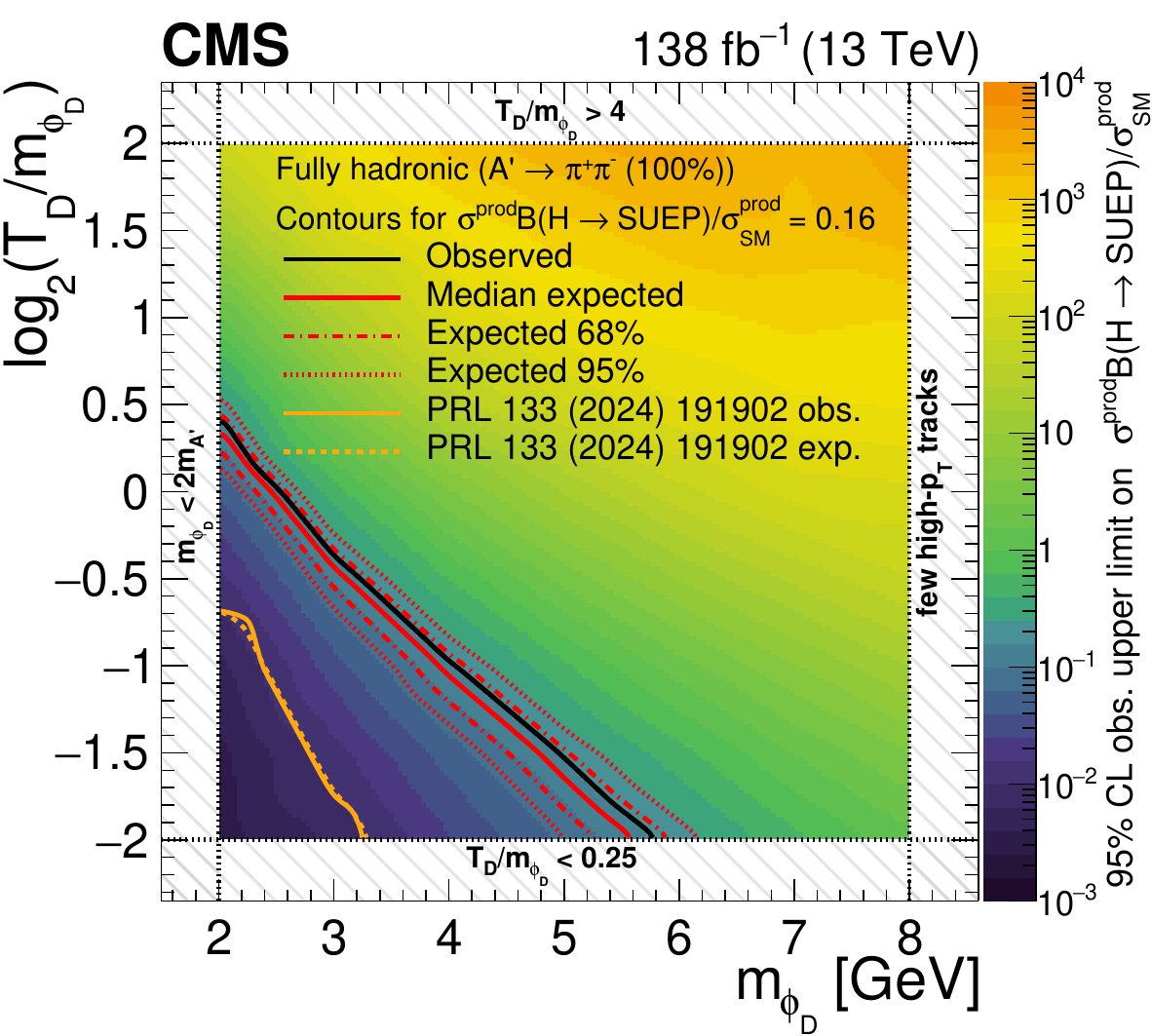}
    \caption{Interpretation of the results for the decay of the 125\GeV Higgs boson mediator to a SUEP, for signal models with $\BR(\PAprime\to\pipi)=100\%$. The observed upper limit from the combination of the \WH and \ZH channels is shown as a function of the signal model parameters \mPhi and $\log_{2}(\TD/\mPhi)$. The contours for each channel are drawn for a benchmark scenario in which the observed or expected upper limit for that channel on \sigmatimesBRoverSM equals 0.16.}
    \label{fig:exclusion_had}
\end{figure}

The results from the combined fit to both channels are shown in Fig.~\ref{fig:fit}. Upper limits at 95\% confidence level (\CL) on \sigmatimesBRoverSM are computed using the \CLs criterion~\cite{Junk:1999kv, Read:2002hq}. The distribution of the likelihood ratio test statistic is estimated using pseudo-experiments and the LHC-style prescription~\cite{CMS:2024onh}. The resulting upper limits on \sigmatimesBRoverSM for several sets of signal models are shown in Figs.~\ref{fig:exclusion},~\ref{fig:exclusion_lep}, and~\ref{fig:exclusion_had}. 

To compare the \VH channel and the gluon fusion channel from the previous CMS result~\cite{CMS:2024nca}, we also show exclusion curves for a benchmark scenario. In the benchmark scenario, $\sigmatimesBRoverSM$ is set to 0.16, which corresponds to the case of $\sigmaProd = \sigmaSM$ and $\BRHSUEP=0.16$, the current best limit on the undetected \PH boson branching fraction from CMS~\cite{CMS:2022dwd}. For the comparison with Ref.~\cite{CMS:2024nca}, a SM production cross section (\sigmaSM) of 45.2\unit{pb} for the gluon fusion production mode is assigned.

No significant deviation from the background prediction is observed. The HEPData entry for this search provides numerical information to facilitate further comparisons~\cite{HEPData}.
This analysis is found to be most sensitive to models with lower \mPhi and \TD, as those tend to produce final states with higher multiplicities of lower-momentum particles. These limits on the production rate of a SUEP shower via an \PH boson improve upon those of Ref.~\cite{CMS:2024nca} by up to two orders of magnitude.

\section{Additional material for interpretation of the results}\label{sec:reinterpretation}

Additional material is provided for the reinterpretation of this analysis in the associated HEPData entry~\cite{HEPData} and the full statistical model repository~\cite{ThisAnalysisMadAnalysis}. An implementation of the analysis in the \MADANALYSIS framework~\cite{Conte:2012fm,Conte:2014zja,Dumont:2014tja,Araz:2020lnp} is provided, along with the full statistical model described in Section~\ref{sec:systematics}, for all signal models considered in the analysis. Additional step-by-step instructions for the processing of alternative signal samples using the \DELPHES~\cite{deFavereau:2013fsa} fast simulation and reconstruction software, event selection using the \MADANALYSIS framework, and statistical inference using the full statistical model are also provided.

Signal samples using the same generator-level parameters and processed with the \DELPHES plus \MADANALYSIS implementation of this search are used to validate this reinterpretation workflow. Relevant distributions in SR-WH and SR-ZH are found to be consistent within 20\% between these samples and those obtained with the full CMS simulation and reconstruction chain detailed in Section~\ref{sec:samples}. The upper limits on \sigmatimesBRoverSM obtained with this alternative setup are consistent within two standard deviations with those obtained with signals processed with the full CMS simulation. Tabulated results of these validation checks can be found in the HEPData record for this analysis~\cite{HEPData}.

\begin{figure}[hbt!]
    \centering
	\includegraphics[width=\cmsFigWidth]{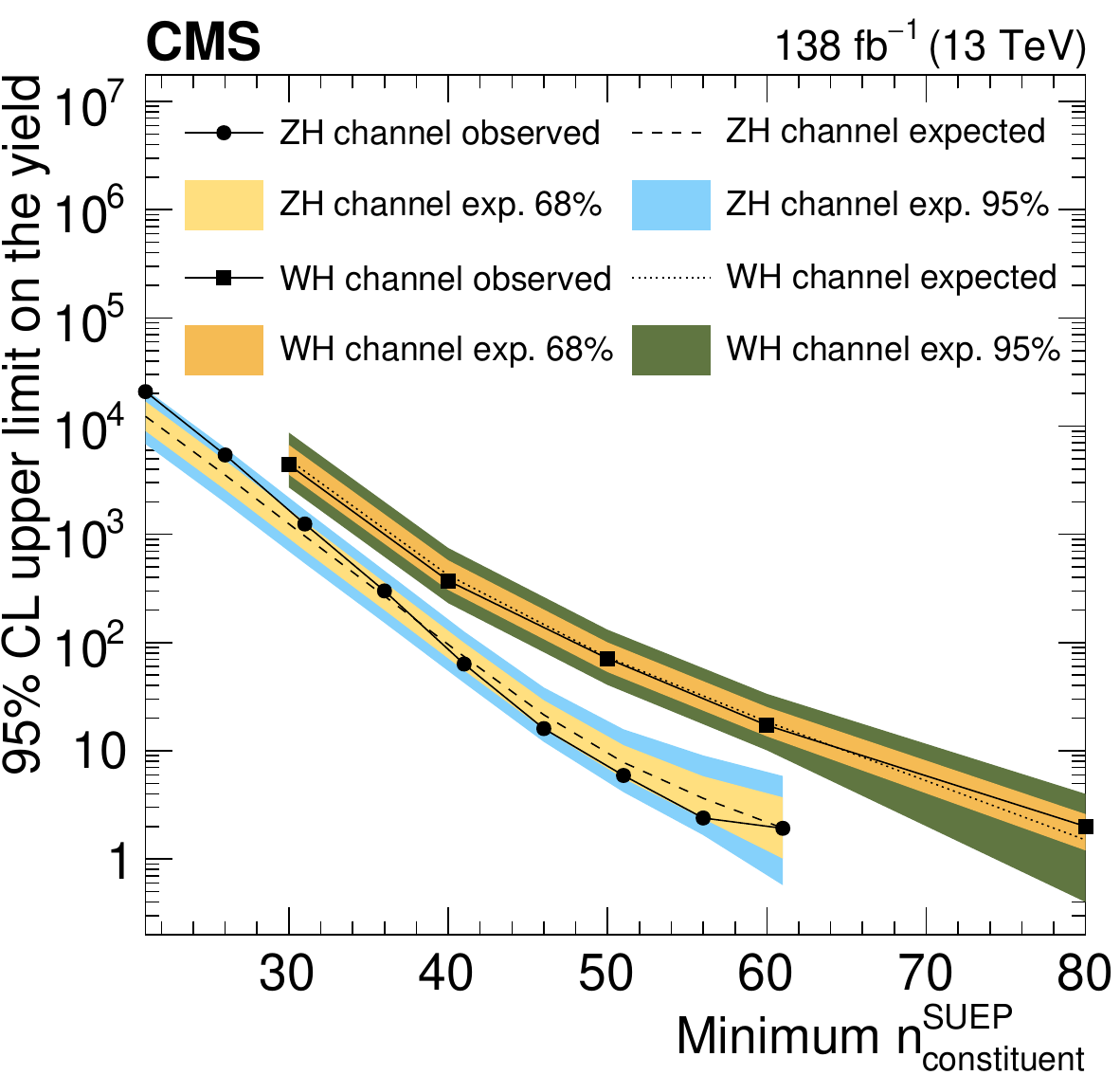}
    \caption{95\% \CL upper limits on the signal yield from the model-agnostic fit for different minimum \nSUEPConstituents values in the \WH and \ZH channels.}
    \label{fig:agnostic}
\end{figure}

Finally, an alternative model-agnostic fit is performed to facilitate reinterpretation of the results without any assumptions about the signal \nSUEPConstituents distribution and to provide simplified, single-bin estimations of the analysis performance. This fit provides upper limits on the signal yields in the extended ABCD SER for different \nSUEPConstituents lower bounds. For each lower bound, all events with higher \nSUEPConstituents values are aggregated into a single bin where the background is estimated using the extended ABCD method. Numerical results are provided separately for the \WH and \ZH channels in the analysis HEPData entry~\cite{HEPData}. Instructions can be found in Ref.~\cite{ThisAnalysisMadAnalysis} to obtain the corresponding predicted yields for other signal samples produced using the \DELPHES plus \MADANALYSIS implementation. By comparing the numbers obtained for these samples with the yield limits above, conclusions can be derived for alternative new physics models without the need to use the full statistical model, but at the cost of reduced sensitivity. The \WH channel provides a weaker limit because of the larger background, but also usually has larger expected signal yield than the \ZH channel (see Fig.~\ref{fig:agnostic}).

\section{Summary}\label{sec:conclusions}
In summary, this Letter presents the first search for final states with a large-radius cluster of low-momentum, spherically distributed particles---a soft unclustered energy pattern or SUEP---and a massive vector boson. Events are selected at trigger level based on the presence of one or two charged leptons arising from the leptonic decays of the \PW or \PZ boson. This strategy improves significantly the sensitivity of the search to SUEP production via a Higgs boson mediator compared to previous results~\cite{CMS:2024nca}, with more significant improvements for signal models with higher multiplicities of low-momentum particles. Additional material is provided for full reinterpretation of the results using \DELPHES and \MADANALYSIS~\cite{ThisAnalysisMadAnalysis}.  

\begin{acknowledgments}
We congratulate our colleagues in the CERN accelerator departments for the excellent performance of the LHC and thank the technical and administrative staffs at CERN and at other CMS institutes for their contributions to the success of the CMS effort. In addition, we gratefully acknowledge the computing centers and personnel of the Worldwide LHC Computing Grid and other centers for delivering so effectively the computing infrastructure essential to our analyses. Finally, we acknowledge the enduring support for the construction and operation of the LHC, the CMS detector, and the supporting computing infrastructure provided by the following funding agencies: SC (Armenia), BMBWF and FWF (Austria); FNRS and FWO (Belgium); CNPq, CAPES, FAPERJ, FAPERGS, and FAPESP (Brazil); MES and BNSF (Bulgaria); CERN; CAS, MoST, and NSFC (China); MINCIENCIAS (Colombia); MSES and CSF (Croatia); RIF (Cyprus); SENESCYT (Ecuador); ERC PRG and PSG, TARISTU24-TK10 and MoER TK202 (Estonia); Academy of Finland, MEC, and HIP (Finland); CEA and CNRS/IN2P3 (France); SRNSF (Georgia); BMFTR, DFG, and HGF (Germany); GSRI (Greece); NKFIH (Hungary); DAE and DST (India); IPM (Iran); SFI (Ireland); INFN (Italy); MSIT and NRF (Republic of Korea); MES (Latvia); LMTLT (Lithuania); MOE and UM (Malaysia); BUAP, CINVESTAV, CONACYT, LNS, SEP, and UASLP-FAI (Mexico); MOS (Montenegro); MBIE (New Zealand); PAEC (Pakistan); MSHE, NSC, and NAWA (Poland); FCT (Portugal); MESTD (Serbia); MICIU/AEI and PCTI (Spain); MOSTR (Sri Lanka); Swiss Funding Agencies (Switzerland); MST (Taipei); MHESI (Thailand); TUBITAK and TENMAK (T\"{u}rkiye); NASU (Ukraine); STFC (United Kingdom); DOE and NSF (USA).

\hyphenation{Rachada-pisek} Individuals have received support from the Marie-Curie program and the European Research Council and Horizon 2020 Grant, contract Nos.\ 675440, 724704, 752730, 758316, 765710, 824093, 101115353, 101002207, 101001205, and COST Action CA16108 (European Union); the Leventis Foundation; the Alfred P.\ Sloan Foundation; the Alexander von Humboldt Foundation; the Science Committee, project no. 22rl-037 (Armenia); the Fonds pour la Formation \`a la Recherche dans l'Industrie et dans l'Agriculture (FRIA) and Fonds voor Wetenschappelijk Onderzoek contract No. 1228724N (Belgium); the Beijing Municipal Science \& Technology Commission, No. Z191100007219010, the Fundamental Research Funds for the Central Universities, the Ministry of Science and Technology of China under Grant No. 2023YFA1605804, the Natural Science Foundation of China under Grant No. 12535004, and USTC Research Funds of the Double First-Class Initiative No.\ YD2030002017 (China); the Ministry of Education, Youth and Sports (MEYS) of the Czech Republic; the Shota Rustaveli National Science Foundation, grant FR-22-985 (Georgia); the Deutsche Forschungsgemeinschaft (DFG), among others, under Germany's Excellence Strategy -- EXC 2121 ``Quantum Universe" -- 390833306, and under project number 400140256 - GRK2497; the Hellenic Foundation for Research and Innovation (HFRI), Project Number 2288 (Greece); the Hungarian Academy of Sciences, the New National Excellence Program - \'UNKP, the NKFIH research grants K 131991, K 133046, K 138136, K 143460, K 143477, K 146913, K 146914, K 147048, 2020-2.2.1-ED-2021-00181, TKP2021-NKTA-64, and 2025-1.1.5-NEMZ\_KI-2025-00004 (Hungary); the Council of Science and Industrial Research, India; ICSC -- National Research Center for High Performance Computing, Big Data and Quantum Computing, FAIR -- Future Artificial Intelligence Research, and CUP I53D23001070006 (Mission 4 Component 1), funded by the NextGenerationEU program, the Italian Ministry of University and Research (MUR) under Bando PRIN 2022 -- CUP I53C24002390006, PRIN PRIMULA 2022RBYK7T (Italy); the Latvian Council of Science; the Ministry of Science and Higher Education, project no. 2022/WK/14, and the National Science Center, contracts Opus 2021/41/B/ST2/01369, 2021/43/B/ST2/01552, 2023/49/B/ST2/03273, and the NAWA contract BPN/PPO/2021/1/00011 (Poland); the Funda\c{c}\~ao para a Ci\^encia e a Tecnologia (Portugal); the National Priorities Research Program by Qatar National Research Fund; MICIU/AEI/10.13039/501100011033, ERDF/EU, ``European Union NextGenerationEU/PRTR", projects PID2022-142604OB-C21, PID2022-139519OB-C21, PID2023-147706NB-I00, PID2023-148896NB-I00, PID2023-146983NB-I00, PID2023-147115NB-I00, PID2023-148418NB-C41, PID2023-148418NB-C42, PID2023-148418NB-C43, PID2023-148418NB-C44, PID2024-158190NB-C22, RYC2021-033305-I, RYC2024-048719-I, CNS2023-144781, CNS2024-154769 and Plan de Ciencia, Tecnolog{\'i}a e Innovaci{\'o}n de Asturias, Spain; the Chulalongkorn Academic into Its 2nd Century Project Advancement Project, the National Science, Research and Innovation Fund program IND\_FF\_68\_369\_2300\_097, and the Program Management Unit for Human Resources \& Institutional Development, Research and Innovation, grant B39G680009 (Thailand); the Eric \& Wendy Schmidt Fund for Strategic Innovation through the CERN Next Generation Triggers project under grant agreement number SIF-2023-004; the Kavli Foundation; the Nvidia Corporation; the SuperMicro Corporation; the Welch Foundation, contract C-1845; and the Weston Havens Foundation (USA).
\end{acknowledgments}

\bibliography{auto_generated}

\cleardoublepage \appendix\section{The CMS Collaboration \label{app:collab}}\begin{sloppypar}\hyphenpenalty=5000\widowpenalty=500\clubpenalty=5000\cmsinstitute{Yerevan Physics Institute, Yerevan, Armenia}
{\tolerance=6000
A.~Hayrapetyan, V.~Makarenko\cmsorcid{0000-0002-8406-8605}, A.~Tumasyan\cmsAuthorMark{1}\cmsorcid{0009-0000-0684-6742}
\par}
\cmsinstitute{Institut f\"{u}r Hochenergiephysik, Vienna, Austria}
{\tolerance=6000
W.~Adam\cmsorcid{0000-0001-9099-4341}, L.~Benato\cmsorcid{0000-0001-5135-7489}, T.~Bergauer\cmsorcid{0000-0002-5786-0293}, M.~Dragicevic\cmsorcid{0000-0003-1967-6783}, P.S.~Hussain\cmsorcid{0000-0002-4825-5278}, M.~Jeitler\cmsAuthorMark{2}\cmsorcid{0000-0002-5141-9560}, N.~Krammer\cmsorcid{0000-0002-0548-0985}, A.~Li\cmsorcid{0000-0002-4547-116X}, D.~Liko\cmsorcid{0000-0002-3380-473X}, M.~Matthewman, J.~Schieck\cmsAuthorMark{2}\cmsorcid{0000-0002-1058-8093}, R.~Sch\"{o}fbeck\cmsAuthorMark{2}\cmsorcid{0000-0002-2332-8784}, M.~Shooshtari\cmsorcid{0009-0004-8882-4887}, M.~Sonawane\cmsorcid{0000-0003-0510-7010}, W.~Waltenberger\cmsorcid{0000-0002-6215-7228}, C.-E.~Wulz\cmsAuthorMark{2}\cmsorcid{0000-0001-9226-5812}
% ...
\par}
\vskip\cmsinstskip
\dag:~Deceased\\
$^{1}$Also at Yerevan State University, Yerevan, Armenia\\
$^{2}$Also at TU Wien, Vienna, Austria\\
% ...
\end{sloppypar}
\end{document}